\documentclass[11pt, a4paper]{article}
\usepackage[T1]{fontenc}
\usepackage{amsfonts}
\usepackage{amsmath}
\usepackage{amssymb}
\usepackage{amsthm}
\usepackage{bbm}
\usepackage{bm}
\usepackage{mathrsfs}
\usepackage{verbatim}
\usepackage{enumitem}
\usepackage{setspace}

\theoremstyle{plain}
\newtheorem{Thm}{Theorem}[section]
\newtheorem{Prop}[Thm]{Proposition}
\newtheorem{Lemma}[Thm]{Lemma}
\newtheorem{Cor}[Thm]{Corollary}

\theoremstyle{definition}

\newtheorem{Remark}[Thm]{Remark}

\newtheorem{Example}[Thm]{Example}

\newtheorem{Assumptions}[Thm]{Assumptions}
\theoremstyle{remark}

\renewenvironment{thebibliography}[1]{%
\begin{oldthebibliography}{#1}%
\setlength{\baselineskip}{.9em}
\small
\setlength{\parskip}{0ex}%
\setlength{\itemsep}{0.1em}%
}%
{%
\end{oldthebibliography}%
}
\newcommand{\q}{\quad}

\newcommand{\eps}{\varepsilon}

\newcommand{\N}{\mathbb{N}}

\newcommand{\R}{\mathbb{R}}

\newcommand{\cF}{\mathcal{F}}

\newcommand{\cA}{\mathcal{A}}

\newcommand{\cE}{\mathcal{E}}

\newcommand{\sC}{\mathscr{C}}
\newcommand{\sD}{\mathscr{D}}

\newcommand{\sJ}{\mathscr{J}}

\newcommand{\sM}{\mathscr{M}}
\newcommand{\sN}{\mathscr{N}}

\newcommand{\fg}{\mathfrak{g}}
\newcommand{\tfg}{\tilde{\fg}}
\newcommand{\fG}{\mathfrak{G}}

\newcommand{\tR}{{\widetilde{R}}}
\newcommand{\tsC}{{\widetilde{\mathscr{C}}}}
\newcommand{\tsN}{{\widetilde{\mathscr{N}}}}

\DeclareMathOperator{\sgn}{sign}
\DeclareMathOperator{\esssup}{ess\, sup}

\DeclareMathOperator{\argmax}{arg\, max}
\newcommand{\hX}{\widehat{X}}
\newcommand{\hY}{\widehat{Y}}
\newcommand{\ty}{\tilde{y}}
\newcommand{\tc}{\tilde{c}}

\newcommand{\cpi}{\bar{\pi}} %
\newcommand{\hpi}{\hat{\pi}}
\newcommand{\hkappa}{\hat{\kappa}}
\newcommand{\ckappa}{\bar{\kappa}} %
\newcommand{\cX}{\bar{X}} %
\newcommand{\tkappa}{\tilde{\kappa}}
\newcommand{\tpi}{\tilde{\pi}}
\newcommand{\tX}{\widetilde{X}}
\newcommand{\sint}{\stackrel{\mbox{\tiny$\bullet$}}{}}
\newcommand{\csC}{\overline{\mathrm{co}}\,(\sC)}

\numberwithin{equation}{section}

\begin{document}%

\title{Power Utility Maximization in Constrained Exponential L\'evy Models\\[1em]
\small{Marcel Nutz\\ETH Zurich, Department of Mathematics, 8092 Zurich, Switzerland\\ \texttt{marcel.nutz@math.ethz.ch} \\
 This Version:  September 1, 2010.}}
\date{}
\maketitle \vspace{-1.5cm}

\begin{abstract}
  We study power utility maximization for exponential L\'evy models with portfolio constraints, where
  utility is obtained from consumption and/or terminal wealth.
  For convex constraints, an explicit solution in terms of the L\'evy triplet is constructed under minimal assumptions
  by solving the Bellman equation. We use a novel transformation of the model to avoid technical conditions. The consequences for $q$-optimal martingale measures are discussed as well as extensions to non-convex constraints.
\end{abstract}

{\small


}\\

\noindent \emph{Acknowledgements.} Financial support by Swiss National Science Foundation Grant PDFM2-120424/1 is gratefully acknowledged. The author thanks Martin Schweizer, Josef Teichmann and two anonymous referees for detailed comments on an earlier version of the manuscript.

\section{Introduction}

A classical problem of mathematical finance is the maximization of expected utility from consumption and/or from terminal wealth for an investor. We consider the special case when the asset prices follow an exponential \emph{L\'evy process} and the investor's preferences are given by a \emph{power utility} function.
This problem was first studied by Merton~\cite{Merton.69} for drifted geometric Brownian motion and
by Mossin~\cite{Mossin.68} and Samuelson~\cite{Samuelson.69} for the discrete-time analogues.
A consistent observation was that when the asset returns
are i.i.d., the optimal portfolio and consumption are given by a constant and a deterministic function, respectively.
This result was subsequently extended to various classes of L\'evy models (see, among others,
Foldes~\cite{Foldes.90},
Framstad et al.~\cite{FramstadOksendalSulem.99},
Benth et al.~\cite{BenthKarlsenReikvam.01, BenthKarlsenReikvam.01b},
Kallsen~\cite{Kallsen.00})  and its general validity was readily conjectured.
We note that the existence of an optimal strategy is known also for much more general models (see Karatzas and \v{Z}itkovi\'c~\cite{KaratzasZitkovic.03}), but \emph{a priori} that strategy is \emph{some} stochastic process without a constructive description.

We prove this conjecture for general L\'evy models under
\emph{minimal} assumptions; in addition, we consider the case where the
choice of the portfolio is constrained to a convex set. The optimal \emph{investment portfolio} is characterized
as the maximizer of a deterministic concave function $\fg$ defined in terms of the L\'evy triplet; and
the maximum of $\fg$ yields the optimal \emph{consumption}. Moreover, the L\'evy triplet characterizes the
finiteness of the value function, i.e., the maximal expected utility.
We also draw the conclusions for the \emph{$q$-optimal equivalent martingale measures} that are linked to utility maximization by convex duality ($q\in (-\infty,1)\setminus\{0\}$); this results in
an explicit existence characterization and a formula for the density process. Finally, some generalizations to non-convex constraints are studied.

Our method consists in solving the \emph{Bellman equation}, which was introduced for general semimartingale
models in Nutz~\cite{Nutz.09b}. In the L\'evy setting, this equation reduces to a Bernoulli ordinary differential equation. There are two main mathematical difficulties.
The first one is to construct the maximizer for $\fg$, i.e., the optimal portfolio.
The necessary compactness is obtained from a minimal no-free-lunch
condition (``no unbounded increasing profit'') via scaling arguments which were developed by Kardaras~\cite{Kardaras.09} for $\log$-utility.
In our setting these arguments require certain integrability properties of the asset returns. Without compromising the generality, integrability is achieved by a linear transformation of the model which replaces the given assets by certain portfolios. We construct the maximizer for $\fg$ in the transformed model and then revert to the original one.

The second difficulty is to \emph{verify} the optimality of the constructed consumption and investment portfolio.
Here we use the general verification theory of~\cite{Nutz.09b} and exploit a well-known property of L\'evy processes,
namely that any L\'evy local martingale is a true martingale.

This paper is organized as follows.
The next section specifies the optimization problem and the notation related to the L\'evy triplet. We also recall the no-free-lunch condition
$\mathrm{NUIP}_{\sC}$ and the opportunity process.
Section~\ref{se:MainResult} states the main result for utility maximization under convex constraints and relates the triplet to
the finiteness of the value function.
The transformation of the model is described in Section~\ref{se:transformation} and the main theorem is proved in
Section~\ref{se:proofMainThm}.
Section~\ref{se:qOptMeas} gives the application to $q$-optimal measures while non-convex constraints are studied in Section~\ref{se:nonconvexConstraints}.
Related literature is discussed in the concluding
Section~\ref{se:literature} as this necessitates technical terminology introduced in the body of the paper.

\section{Preliminaries}

The following notation is used. If $x,y\in\R$ are reals, $x^+=\max\{x,0\}$ and $x\wedge y=\min\{x,y\}$. We set $1/0:=\infty$ where necessary. If $z\in\R^d$ is a $d$-dimensional vector, $z^i$ is its $i$th coordinate and $z^\top$ its transpose. Given $A\subseteq \R^d$,
$A^\bot$ denotes the Euclidean orthogonal complement and $A$ is said to be \emph{star-shaped} (with respect to the origin) if $\lambda A\subseteq A$ for all $\lambda \in [0,1]$.  If $X$ is an $\R^d$-valued semimartingale
and $\pi\in L(X)$ is an $\R^d$-valued predictable integrand, the vector stochastic integral is a scalar semimartingale with initial value zero and denoted by $\int \pi\,dX$ or by $\pi\sint X$. Relations between measurable functions
hold almost everywhere unless otherwise stated. Our reference for any unexplained notion or notation from stochastic calculus is Jacod and Shiryaev~\cite{JacodShiryaev.03}.

\subsection{The Optimization Problem}

 We fix the time horizon $T\in(0,\infty)$ and a probability space $(\Omega,\cF,P)$ with a filtration $(\cF_t)_{t\in[0,T]}$  satisfying the usual assumptions of right-continuity and completeness, as well as $\cF_0=\{\emptyset,\Omega\}$ $P$-a.s.
We consider an $\R^d$-valued \emph{L\'evy process} $R=(R^1,\dots,R^d)$ with $R_0=0$.
That is, $R$ is a c\`adl\`ag semimartingale with stationary independent increments as defined in~\cite[II.4.1(b)]{JacodShiryaev.03}.
It is not relevant for us whether $R$ generates the filtration.
The stochastic exponential $S=\cE(R)=(\cE(R^1),\dots,\cE(R^d))$ represents the discounted price processes of $d$ risky assets,
while $R$ stands for their returns. If one wants to model only positive prices, one can equivalently use the ordinary exponential
(see, e.g., \cite[Lemma~4.2]{Kallsen.00}). Our agent also has a bank account paying zero interest at his disposal.

The agent is endowed with a deterministic initial capital $x_0>0$. A \emph{trading strategy} is a predictable $R$-integrable $\R^d$-valued process $\pi$, where the $i$th component is interpreted as the fraction of wealth (or the portfolio proportion) invested in the $i$th risky asset.

We want to consider two cases. Either consumption occurs only at the terminal time $T$ (utility from ``terminal wealth'' only); or there is intermediate consumption plus a bulk consumption at the time horizon. To unify the notation, we define
 \[
   \delta:=
  \begin{cases}
    1 & \text{in the case with intermediate consumption}, \\
    0 & \text{otherwise}.
  \end{cases}
 \]
It will be convenient to parametrize the consumption strategies as a fraction of the current wealth. A \emph{propensity to consume} is a nonnegative optional process $\kappa$ satisfying $\int_0^T \kappa_s\,ds<\infty$ $P$-a.s.
The \emph{wealth process} $X(\pi,\kappa)$ corresponding to a pair $(\pi,\kappa)$ is defined by the stochastic exponential
\begin{equation*}%
   X(\pi,\kappa)=x_0\cE\big(\pi\sint R - \delta \textstyle{\int} \kappa_s\,ds\big).
\end{equation*}
In particular, the number of shares of the $i$th asset held in the portfolio is then given by $\pi^i X(\pi,\kappa)_- / S^i_-$ (on the set $\{S^i_-\neq0\}$).

Let $\sC\subseteq\R^d$ be a set containing the origin; we refer to $\sC$ as ``the constraints''.
The set of (constrained) \emph{admissible} strategies is
\[
  \cA(x_0):=\big\{(\pi,\kappa): X(\pi,\kappa)>0\mbox{ and }\pi_t(\omega)\in\sC\;\mbox{ for all }(\omega,t)\in\Omega\times[0,T]\big\}.
\]
We fix the initial capital $x_0$ and usually write $\cA$ for $\cA(x_0)$.
Given $(\pi,\kappa)\in\cA$, $c:=\kappa X(\pi,\kappa)$ is the corresponding \emph{consumption rate}
and $X=X(\pi,\kappa)$ satisfies the self-financing condition $X_t=x_0+\int_0^t  X_{s-} \pi_s\,dR_s-\delta \int_0^t c_s\,ds$.

Let $p\in (-\infty,0)\cup(0,1)$. We use the power utility function
\begin{equation*}%
  U(x):=\tfrac{1}{p}x^p,\q x\in (0,\infty)
\end{equation*}
to model the preferences of the agent. The constant $\beta:=(1-p)^{-1}>0$ is the relative risk tolerance of~$U$. Note that we exclude the logarithmic utility,
which corresponds to $p=0$ and for which the optimal strategies are known explicitly even for semimartingale models (see Karatzas and Kardaras~\cite{KaratzasKardaras.07}).

Let $(\pi,\kappa)\in\cA$ and $X=X(\pi,\kappa)$, $c=\kappa X$.
The corresponding \emph{expected utility} is
$E\big[\delta\int_0^T U(c_t)\,dt + U(X_T)\big]$.
The \emph{value function} is given by
\begin{equation*}%
  u(x_0):=\sup_{\cA(x_0)}E\Big[\delta\int_0^T U(c_t)\,dt + U(X_T)\Big],
\end{equation*}
where the supremum is taken over all $(c,X)$ which correspond to some $(\pi,\kappa)\in\cA(x_0)$.
We say that the utility maximization problem is \emph{finite} if $u(x_0)<\infty$;
this always holds if $p<0$ as then $U<0$.
If $u(x_0)<\infty$, $(\pi,\kappa)$ is \emph{optimal} if the corresponding $(c,X)$ satisfy
$E\big[\delta\int_0^T U(c_t)\,dt + U(X_T)\big]=u(x_0)$.

\subsection{L\'evy Triplet, Constraints, No-Free-Lunch Condition}\label{se:triplet}

Let $(b^R,c^R,F^R)$ be the L\'evy triplet of $R$ with respect to some fixed
cut-off function $h: \R^d\to\R^d$ (i.e., $h$ is bounded and  $h(x)=x$ in a neighborhood of $x=0$).
This means that $b^R\in\R^d$, $c^R\in\R^{d\times d}$ is a nonnegative definite matrix, and $F^R$ is a L\'evy measure on $\R^d$, i.e., $F^R\{0\}=0$ and
\begin{equation}\label{eq:LevyIntegrates}
  \int_{\R^d} 1\wedge |x|^2 \,F^R(dx)<\infty.
\end{equation}
The process $R$ can be represented as
\[
  R_t=b^Rt + R_t^c + h(x)\ast (\mu^R_t-\nu^R_t) + (x-h(x))\ast \mu^R_t.
\]
Here $\mu^R$ is the integer-valued random measure associated with the jumps of $R$
and $\nu^R_t=tF^R$ is its compensator. Moreover, $R^c$ is the continuous martingale part, in fact,
$R_t^c=\sigma W_t$, where $\sigma\in\R^{d\times d}$ satisfies $\sigma\sigma^\top=c^R$ and $W$ is a $d$-dimensional standard Brownian motion. We refer to~\cite[II.4]{JacodShiryaev.03} or Sato~\cite{Sato.99} for background material concerning L\'evy processes.

We introduce some subsets of $\R^d$ to be used in the sequel; the terminology follows~\cite{Kardaras.09}. The first two are related to the ``budget constraint'' $X(\pi,\kappa)>0$.
The \emph{natural constraints} are given by
\[
  \sC^0:=\Big\{y\in\R^d:\, F^R\big[x\in\R^d:\, y^\top x < -1\big]=0\Big\};
\]
clearly $\sC^0$ is closed, convex, and contains the origin.
We also consider the slightly smaller set
\[
  \sC^{0,*}:=\Big\{y\in\R^d:\, F^R\big[x\in\R^d:\, y^\top x \leq -1\big]=0\Big\}.
\]
It is convex, contains the origin, and its closure equals $\sC^0$, but it is a proper subset in general.
The meaning of these sets is explained by

\begin{Lemma}\label{le:positivity}
  A process $\pi\in L(R)$ satisfies $\cE(\pi\sint R) \geq 0\, (>0)$ if and only if $\pi$ takes values in  $\sC^0\,(\sC^{0,*})$ $P\otimes dt$-a.e.
\end{Lemma}

See, e.g.,~\cite[Lemma~2.5]{Nutz.09b} for the proof. We observe that $\sC^{0,*}$ corresponds to the admissible portfolios; however,
the set $\sC^0$ also turns out to be useful, due to the closedness.

The linear space of \emph{null-investments} is defined by
\[
  \sN:=\big\{y\in\R^d:\,y^\top b^R=0,\,y^\top c^R=0,\,F^R[x:\,y^\top x\neq 0]=0 \big\}.
\]
Then $H\in L(R)$ satisfies $H\sint R\equiv0$ if and only if $H$ takes values in $\sN$ $P\otimes dt$-a.e.
In particular, two portfolios $\pi$ and $\pi'$ generate the same wealth process if and only if $\pi-\pi'$ is $\sN$-valued.

We recall the set $0\in \sC\subseteq\R^d$ of portfolio constraints.
The set $\sJ\subseteq\R^d$ of \emph{immediate arbitrage opportunities} is defined by
\begin{equation*}%
  \sJ=\Big\{y:\, y^\top c^R =0,\; F^R[y^\top x<0]=0,\; y^\top b^R -\int y^\top h(x)\,F^R(dx)\geq0\Big\}\setminus \sN.
\end{equation*}
Note that for $y\in\sJ$, the process $y^\top R$ is increasing and nonconstant.
For a subset $G$ of $\R^d$, the recession cone is given by $\check{G}:=\bigcap_{a\geq0} aG$.
Now the condition $\mathrm{NUIP}_{\sC}$ (no unbounded increasing profit) can be defined by
\[
  \mathrm{NUIP}_{\sC}\q \Longleftrightarrow\q \sJ \cap \check{\sC} = \emptyset
\]
(cf.~\cite[Theorem~4.5]{Kardaras.09}). This is equivalent to
$\sJ\cap (\sC\cap\sC^0)^{ \check{} }=\emptyset$ because $\sJ\subseteq \check{\sC}^0$, and it
means that if a strategy leads to an increasing nonconstant wealth process, then that strategy cannot be scaled arbitrarily.
This is a minimal no-free lunch condition (cf.\ Remark~\ref{rk:onAssumptions}(c)); we refer to~\cite{Kardaras.09} for more information about free lunches in
exponential L\'evy models.
We give a simple example to illustrate the objects.

\begin{Example}\label{ex:simpleIllustration}
  Assume there is only one asset ($d=1$), that its price is strictly positive, and that it can jump arbitrarily close to zero and arbitrarily high. In formulas,
  $F^R(-\infty,-1]=0$ and for all $\eps>0$, $F^R(-1,-1+\eps]>0$ and $F^R[\eps^{-1},\infty)>0$.

  Then $\sC^0=\sC^{0,*}=[0,1]$ and $\sN=\{0\}$. In this situation $\mathrm{NUIP}_{\sC}$ is satisfied for any set $\sC$, both because
  $\sJ=\emptyset$ and because $\check{\sC^0}=\{0\}$.
  If the price process is merely nonnegative and $F^R\{-1\}>0$, then $\sC^{0,*}=[0,1)$
  while the rest stays the same.
\end{Example}
In fact, most of the scalar models presented in Schoutens~\cite{Schoutens.03} correspond to the first part of Example~\ref{ex:simpleIllustration} for nondegenerate choices of the parameters.

\subsection{Opportunity Process}

Assume $u(x_0)<\infty$ and let $(\pi,\kappa)\in\cA$. For fixed $t\in[0,T]$, the set of ``compatible'' controls is
$\cA(\pi,\kappa,t):=\big\{(\tilde{\pi},\tilde{\kappa})\in\cA:\, (\tilde{\pi},\tilde{\kappa})=(\pi,\kappa)\mbox{ on }[0,t]\big\}$.
By~\cite[Proposition~3.1, Remark~3.7]{Nutz.09a} there exists a unique c\`adl\`ag semimartingale $L$, called \emph{opportunity process}, such that
\begin{equation*}%
    L_t\,\tfrac{1}{p}\big(X_t(\pi,\kappa)\big)^p
       = \mathop{\esssup}_{\cA(\pi,\kappa,t)} E\Big[\delta\int_t^T U(\tc_s)\,ds + U(\tilde{X}_T)\Big|\cF_t\Big],
\end{equation*}
where the supremum is taken over all consumption and wealth pairs $(\tc,\tX)$ corresponding to some $(\tpi,\tkappa)\in\cA(\pi,\kappa,t)$.
The right hand side above is known as the \emph{value process} of our control problem and its factorization appearing on the left hand side is a consequence of the $p$-homogeneity of $U$. We shall see that in the present (Markovian) L\'evy setting, the opportunity process is simply a deterministic function. In particular, the value process coincides with the more familiar ``dynamic value function'' in the sense of Markovian dynamic programming, $u_t(x)= L_t\,\tfrac{1}{p} x^p$, evaluated at $X_t(\pi,\kappa)$.

\section{Main Result}\label{se:MainResult}

We can now formulate the main theorem for the convex-constrained case;
the proofs are given in the two subsequent sections. We consider the following conditions (not to be understood as standing assumptions).

\begin{Assumptions}\label{as:convex}\hspace{1em}
 \begin{enumerate}[topsep=3pt, partopsep=0pt, itemsep=1pt,parsep=2pt]
   \item $\sC$ is convex.
   \item The orthogonal projection of $\sC\cap\sC^0$ onto $\sN^\bot$ is closed.
   \item $\mathrm{NUIP}_{\sC}$ holds.
   \item $u(x_0)<\infty$, i.e., the utility maximization problem is finite.
 \end{enumerate}
\end{Assumptions}

To state the result, we define for $y\in\sC^0$ the deterministic function
\begin{equation}\label{eq:frakg}
   \fg(y) :=y^\top b^R + \tfrac{(p-1)}{2} y^\top c^R y
            + \int_{\R^d} \big\{p^{-1}(1+y^\top x)^p - p^{-1} - y^\top h(x)\big\}\, F^R(dx).
 \end{equation}
As we will see later, this concave function is well defined with values in $\R\cup \{\sgn(p) \infty\}$.

\begin{Thm}\label{th:MainResultConvex}
  Under Assumptions~\ref{as:convex}, there exists an optimal strategy $(\hpi,\hkappa)$ such that
  $\hpi$ is a constant vector and $\hkappa$ is deterministic. Here $\hpi$ is characterized by
  \[
      \hpi \in \argmax_{\sC\cap\sC^0} \fg
  \]
  and, in the case with intermediate consumption,
  \[
    \hkappa_t=a\big((1+a)e^{a (T-t)} - 1\big)^{-1},
  \]
  where $a:=\tfrac{p}{1-p}\max_{\sC\cap\sC^0} \fg$.
  The opportunity process is given by
  \[
  L_t=\begin{cases}
    \exp\big(a(1-p)(T-t)\big) & \text{without intermediate consumption}, \\
    a^{p-1}\big[(1+a)e^{a (T-t)} - 1\big]^{1-p} & \text{with intermediate consumption}.
  \end{cases}
  \]
\end{Thm}

\begin{Remark}[{\cite[Remark~3.3]{Nutz.09b}}]
  The propensity to consume $\hkappa$ is unique.
  The optimal portfolio $\hpi$ and $\argmax_{\sC\cap\sC^0} \fg$ are unique modulo $\sN$; i.e., if $\pi^*$
  is another optimal portfolio (or maximizer), then $\hpi-\pi^*$ takes values in $\sN$.
  Equivalently, the wealth processes coincide.
\end{Remark}

We comment on Assumptions~\ref{as:convex}.

\begin{Remark}\label{rk:onAssumptions}
 (a)~~Convexity of $\sC$ is of course not necessary to have a solution. We give some generalizations
 in Section~\ref{se:nonconvexConstraints}.

 (b)~~The projection of $\sC\cap\sC^0$ onto $\sN^\bot$ should be understood as ``effective domain'' of the optimization problem. Without the closedness in (ii), there are examples with non-existence of an optimal strategy even for drifted Brownian motion and closed convex cone constraints; see Example~\ref{ex:coneNotClosed} below. One can note that closedness of $\sC$ implies (ii) if $\sN\subseteq \sC$ and $\sC$ is convex (as this implies $\sC=\sC+\sN$, see \cite[Remark~2.4]{Kardaras.09}). Similarly, (ii) holds whenever the projection of $\sC$ to $\sN^\bot$ is closed: if $\Pi$ denotes the projector, $\sC^0=\sC^0+\sN$ yields $\Pi (\sC\cap\sC^0)=(\Pi \sC)\cap\sC^0$ and $\sC^0$ is closed. This includes the cases where $\sC$ is closed and polyhedral, or compact.

 (c)~~Suppose that $\mathrm{NUIP}_{\sC}$ does not hold. If $p\in(0,1)$, it is obvious that $u(x_0)=\infty$. If $p<0$, there exists no optimal strategy, essentially because adding a suitable arbitrage strategy would always yield a higher expected utility.
 See~\cite[Proposition~4.19]{KaratzasKardaras.07} for a proof.

 (d)~~If $u(x_0)=\infty$, either there is no optimal strategy, or there
 are infinitely many strategies yielding infinite expected utility. It would be inconvenient to call the latter optimal. Indeed, using that $u(x_0/2)=\infty$, one can typically construct such strategies which also exhibit intuitively suboptimal behavior (such as throwing away money by a ``suicide strategy''; see~Harrison and Pliska~\cite[\S 6.1]{HarrisonPliska.81}). Hence we require (iv) to have a meaningful solution to our problem---the relevant question is how to characterize this condition in terms of the model.
\end{Remark}

The following example illustrates how non-existence of an optimal portfolio may occur when Assumption~\ref{as:convex}(ii) is violated. It is based on Czichowsky et al.~\cite[\S2.2]{CzichowskyWestrayZheng.08}, where the authors give
an example of a set of stochastic integrals which is not closed in the semimartingale topology.
We denote by $e_j$, $1\leq j\leq d$ the unit vectors in $\R^d$, i.e., $e^i_j=\delta_{ij}$.

\begin{Example}[$\delta=0$]\label{ex:coneNotClosed}
   Let $W$ be a standard Brownian motion in $\R^3$ and
   \[
      \sigma = \left(
               \begin{matrix}
                 1 & 0  & 0 \\
                 0 & 1 & -1 \\
                 0 & -1 & 1 \\
               \end{matrix}
             \right);\quad \sC=\Big\{y\in\R^3:\, \big|y^1\big|^2+\big|y^2\big|^2\leq \big|y^3\big|^2,\, y^3\geq0 \Big\}.
   \]
   Let $R_t=bt+\sigma W_t$, where $b:=e_1$ is orthogonal to $\ker \sigma^\top= \R(0,1,1)^\top$. Thus
   $\sN=\ker \sigma^\top$ and $\sN^\bot$ is spanned by $e_1$ and $e_2-e_3$. The closed convex cone $\sC$ is ``leaning'' against the plane $\sN^\bot$ and the orthogonal projection of $\sC$ onto $\sN^\bot$ is an open half-plane plus the origin. The vectors $\alpha e_1$ with $\alpha\in\R\setminus\{0\}$
   are not contained in this half-plane but in its closure.

   The optimal portfolio $\hpi$ for the \emph{unconstrained} problem lies on this boundary.
   Indeed, $\mathrm{NUIP}_{\R^3}$ holds and Theorem~\ref{th:MainResultConvex} yields $\hpi= \beta (\sigma\sigma^\top)^{-1} e_1=\beta e_1$, where $\beta=(1-p)^{-1}$. This is simply Merton's optimal portfolio in the market consisting only of the first asset.
   By construction we find vectors
   $\pi^n\in\sC$ whose projections to $\sN^\bot$ converge to $\hpi$ and it is easy to see that $E[U(X_T(\pi^n))]\to E[U(X_T(\hpi))]$.
   Hence the value functions for the constrained and the unconstrained problem are identical.
   Since the solution $\hpi$ of the unconstrained problem is unique modulo $\sN$, this implies that if the
   constrained problem has a solution, it has to agree with $\hpi$, modulo $\sN$. But $(\{\hpi\}+\sN)\cap \sC=\emptyset$,
   so there is no solution.
\end{Example}

The rest of the section is devoted to the characterization of Assumption~\ref{as:convex}(iv) by the jump characteristic $F^R$ and the set $\sC$;
this is intimately related to the moment condition
\begin{equation}\label{eq:LevyPthMoment}
  \int_{\{|x|>1\}} |x|^p \,F^R(dx)<\infty.
\end{equation}

We start with a partial result; again $e_j$, $1\leq j\leq d$ denote the unit vectors.

\begin{Prop}\label{pr:levyFinitenessPrelim}
  Let $p\in (0,1)$.
  \begin{enumerate}[topsep=3pt, partopsep=0pt, itemsep=1pt,parsep=2pt]
    \item  Under Assumptions~\ref{as:convex}(i)-(iii), \eqref{eq:LevyPthMoment} implies $u(x_0)<\infty$.
    \item  If $e_j\in \sC\cap\sC^{0,*}$ for all $1\leq j \leq d$, then $u(x_0)<\infty$ implies~\eqref{eq:LevyPthMoment}.
  \end{enumerate}
\end{Prop}

By Lemma~\ref{le:positivity} the $j$th asset has a positive price if and only if $e_j\in\sC^{0,*}$.
Hence we have the following consequence of Proposition~\ref{pr:levyFinitenessPrelim}.

\begin{Cor}
   In an unconstrained exponential L\'evy model with positive asset prices satisfying $\mathrm{NUIP}_{\R^d}$,
   $u(x_0)<\infty$ is equivalent to \eqref{eq:LevyPthMoment}.
\end{Cor}

The implication $u(x_0)<\infty \Rightarrow\eqref{eq:LevyPthMoment}$ is essentially true also in the general case;
more precisely, it holds in an equivalent model. As a motivation, consider the case where
either $\sC=\{0\}$ or $\sC^0=\{0\}$. The latter occurs, e.g., if $d=1$ and the asset has
jumps which are unbounded in both directions. Then the statement $u(x_0)<\infty$ carries no
information about $R$ because $\pi\equiv0$ is the only admissible portfolio. On the other hand, we are
not interested in assets that cannot be traded, and may as well remove them
from the model. This is part of the following result.

\begin{Prop}\label{pr:levyFiniteness}
  There exists a linear transformation $(\tR,\tsC)$ of the model $(R,\sC)$, which is equivalent in that it
  admits the same wealth processes, and has the following properties:
  \begin{enumerate}[topsep=3pt, partopsep=0pt, itemsep=1pt,parsep=2pt]
  \item the prices are strictly positive,
  \item the wealth can be invested in each asset (i.e., $\pi\equiv e_j$ is admissible),
  \item if $u(x_0)<\infty$ holds for $(R,\sC)$, it holds also in the model $(\tR,\tsC)$
        and $\int_{\{|x|>1\}} |x|^p \,F^{\tR}(dx)<\infty$.
  \end{enumerate}
\end{Prop}

The details of the construction are given in the next section, where we also show that Assumptions~\ref{as:convex}
carry over to $(\tR,\tsC)$.

\section{Transformation of the Model}\label{se:transformation}

This section contains the announced linear transformation
of the market model. Assumptions~\ref{as:convex} are not used.
We first describe how any linear transformation affects our objects.

\begin{Lemma}\label{le:linTransform}
  Let $\Lambda$ be a $d\times d$-matrix and define $\tR:=\Lambda R$.
  Then $\tR$ is a L\'evy process with triplet $b^{\tR}=\Lambda b^R$, $c^\tR=\Lambda c^R \Lambda^\top$ and $F^\tR(\cdot)=F^R(\Lambda^{-1} \cdot)$. Moreover, the corresponding natural constraints and null-investments are given by $\tsC^0:=(\Lambda^T)^{-1}\sC^0$ and
  $\tsN:=(\Lambda^T)^{-1}\sN$ and the corresponding function $\tfg$ satisfies $\tfg(z)=\fg(\Lambda^\top z)$.
\end{Lemma}
The proof is straightforward and omitted.
Of course, $\Lambda^{-1}$ refers to the preimage if $\Lambda$ is not invertible.
Given $\Lambda$, we keep the notation from Lemma~\ref{le:linTransform} and introduce also $\tsC:=(\Lambda^T)^{-1}\sC$ as well as $\tsC^{0,*}:=(\Lambda^T)^{-1}\sC^{0,*}$.

\begin{Thm}\label{th:levyChangeCoords}
  There exists a $d\times d$-matrix $\Lambda$ such that for all $1\leq j\leq d$,
  \begin{enumerate}[topsep=3pt, partopsep=0pt, itemsep=1pt,parsep=2pt]
  \item $\Delta \tR^j> -1$ up to evanescence,
  \item $e_j\in\tsC\cap\tsC^{0,*}$,
  \item the model $(\tR,\tsC)$ admits the same wealth processes as $(R,\sC)$.
  \end{enumerate}
\end{Thm}

\begin{proof}
   We treat the components one by one. Pick any vector $y_1\in \sC\cap\sC^{0,*}$ such that $y_1^1\neq0$; if there is no such vector, set $y_1=0$. We replace the first asset $R^1$ by the process $y_1^\top R$. In other words, we replace $R$ by $\Lambda_1 R$, where $\Lambda_1$ is the matrix
   \[
   \Lambda_1 = \left(
               \begin{matrix}
                 y_1^1 & y_1^2  & \cdots  & y_1^d \\
                   & 1 &   &   \\
                   &   & \ddots &   \\
                   &   &   &   1\\
               \end{matrix}
             \right).
   \]
   The new natural constraints are $(\Lambda_1^\top)^{-1} \sC^0$ and we replace $\sC$ by $(\Lambda_1^{\top})^{-1} \sC$.
   Note that $e_1\in (\Lambda_1^\top)^{-1} (\sC\cap\sC^{0,*})$ because $\Lambda_1^\top e_1=y_1\in \sC\cap\sC^{0,*}$ by construction. Similarly, $(\Lambda_1^\top \psi) \sint R= \psi \sint (\Lambda_1 R)$ whenever $\Lambda_1^\top \psi\in L(R)$.
   Therefore, to show that the new model admits the same wealth processes as the old one, we have to show that
   for every $\sC\cap\sC^{0,*}$-valued process $\pi\in L(R)$ there exists a predictable $\psi$ such that
   $\Lambda_1^\top \psi=\pi$; note that this already implies $\psi\in L(\Lambda_1 R)$ and that $\psi$ takes values in $(\Lambda_1^{\top})^{-1}(\sC\cap\sC^{0,*})$. If $\Lambda_1^\top$ is invertible, we take $\psi:=(\Lambda_1^\top)^{-1}\pi$. Otherwise $\pi^1\equiv 0$ by construction and we choose $\psi^1\equiv0$ and $\psi^j=\pi^j$ for $j\geq2$; this is the same as inverting $\Lambda_1^\top$ on its image.

   We proceed with the second component of the new model in the same way, and then continue until the last one.
   We obtain matrices $\Lambda_j$, $1\leq j\leq d$ and set $\Lambda=\Lambda_d\cdots\Lambda_1$.
   The construction and $\Lambda_i^\top e_j=e_j$ for $i\neq j$ imply $e_j \in (\Lambda^\top)^{-1} (\sC\cap\sC^{0,*})$,
   which is (ii), and (i) is a consequence of (ii).
\end{proof}

From now on let $\Lambda$ and $\tR$ be as in Theorem~\ref{th:levyChangeCoords}.

\begin{Cor}\label{co:levyChangeCoords}
  \begin{enumerate}[topsep=3pt, partopsep=0pt, itemsep=1pt,parsep=2pt]
  \item The value functions for $(R,\sC)$ and $(\tR,\tsC)$ coincide.
  \item The opportunity processes for $(R,\sC)$ and $(\tR,\tsC)$ coincide.
  \item $\sup_{\tsC\cap\tsC^{0,*}} \tfg = \sup_{\sC\cap\sC^{0,*}}\fg$.
  \item $z\in \argmax_{\tsC\cap\tsC^{0,*}} \tfg$\; if and only if \;$\Lambda^\top z\in \argmax_{\sC\cap\sC^{0,*}} \fg$.
  \item $(\pi,\kappa)$ is an optimal strategy for $(\tR,\tsC)$ if and only if $(\Lambda^\top \pi,\kappa)$ is optimal for
  $(R,\sC)$.
  \item $\mathrm{NUIP}_{\tsC}$ holds for $\tR$ if and only if $\mathrm{NUIP}_{\sC}$ holds for $R$.
  \end{enumerate}
\end{Cor}

\begin{proof}
  This follows from Theorem~\ref{th:levyChangeCoords}(iii) and Lemma~\ref{le:linTransform}.
\end{proof}

The transformation also preserves certain properties of the constraints.

\begin{Remark}\label{rk:shapePreservation}
  (a)~~If $\sC$ is closed (star-shaped, convex), then $\tsC$ is also closed (star-shaped, convex).

  (b)~~Let $\sC$ be compact, then $\tsC$ is compact only if $\Lambda$ is invertible. However,
  the relevant properties for Theorem~\ref{th:levyChangeCoords} are that $\Lambda^\top \tsC=\sC\cap (\Lambda^\top\R^d)$ and that $e_j\in\tsC$ for $1\leq j\leq d$;
  we can equivalently substitute $\tsC$ by a compact set having these properties.
  If $\Lambda^\top$ is considered as a mapping $\R^d\to \Lambda^\top \R^d$ , it admits a continuous right-inverse $f$,
  and $(\Lambda^\top)^{-1} \sC = f(\sC\cap \Lambda^\top\R^d)+\ker(\Lambda^\top)$. Here $f(\sC\cap \Lambda^\top\R^d)$ is compact and contained in $B_r=\{x\in\R^d:\, |x|\leq r\}$ for some $r\geq 1$. The set
  $\widehat{\sC}:= \big[f(\sC\cap \Lambda^\top\R^d)+\ker(\Lambda^\top)\big]\cap B_r$ has the two desired properties.
\end{Remark}

Next, we deal with the projection of $\tsC\cap\tsC^0$ onto $\tsN^\bot$.
We begin with a ``coordinate-free'' description for its closedness; it can be seen as a simple static version of the main result in Czichowsky and Schweizer~\cite{CzichowskySchweizer.09a}.

\begin{Lemma}\label{le:closednessProbab}
  Let $\sD\subseteq \R^d$ be a nonempty set and let $\sD'$ be its orthogonal projection onto $\sN^\bot$.
  Then $\sD'$ is closed in $\R^d$ if and only if $\{y^\top R_T:\, y\in \sD\}$ is closed for convergence in probability.
\end{Lemma}

\begin{proof}
  Recalling the definition of $\sN$, we may assume that $\sD=\sD'$.
  If $(y_n)$ is a sequence in $\sD$ with some limit $y_*$, clearly $y_n^\top R_T\to y_*^\top R_T$ in probability.
  If $\{y^\top R_T:\, y\in \sD\}$ is closed, it follows that $y_*\in\sD$ because $\sD\cap\sN=\{0\}$; hence $\sD$ is closed.

  Conversely, let $y_n\in\sD$ and assume $y_n^\top R_T\to Y$ in probability for some $Y\in L^0(\cF)$. With $\sD\cap\sN=\{0\}$ it follows that $(y_n)$ is bounded, therefore, it has a subsequence which converges to some $y_*$. If $\sD$ is closed, then $y_*\in\sD$ and we
  conclude that $Y=y_*^\top R_T$, showing closedness in probability.
\end{proof}

\begin{Lemma}
  Assume that $\sC\cap\sC^{0,*}$ is dense in $\sC\cap\sC^0$. Then
  the orthogonal projection of $\sC\cap\sC^0$ onto $\sN^\bot$ is closed if and only if this holds for $\tsC\cap\tsC^0$ and $\tsN^\bot$.
\end{Lemma}

\begin{proof}

  (i)~~Recall the construction of $\Lambda=\Lambda_d\cdots\Lambda_1$ from the proof of Theorem~\ref{th:levyChangeCoords}.
  Assume first that $\Lambda=\Lambda_i$ for some $1\leq i\leq d$. In a first step,
  we show
  \begin{equation}\label{eq:proofProjClosed}
    \Lambda^\top (\tsC\cap\tsC^0) =  \sC\cap\sC^0.
  \end{equation}
  By construction, either $\Lambda$ is invertible, in which case the claim is clear, or otherwise $\Lambda^\top$ is the orthogonal projection of $\R^d$ onto the hyperplane $H_i=\{y\in\R^d:\, y^i=0\}$ and $\sC\cap\sC^{0,*}\subseteq H_i$. By the density assumption it follows that $\sC\cap\sC^0\subseteq H_i$. Thus
  $(\Lambda^\top)^{-1} (\sC\cap\sC^0) = \tsC\cap\tsC^0 + H_i^\bot$, the sum being orthogonal, and
  $\Lambda^\top [(\Lambda^\top)^{-1} (\sC\cap\sC^0)]=\sC\cap\sC^0$ as claimed. We also note that
  \begin{equation}\label{eq:proofProjClosed2}
    (\Lambda^\top)^{-1} (\sC\cap\sC^{0,*}) \subseteq (\Lambda^\top)^{-1} (\sC\cap\sC^0) \quad\mbox{is dense.}
  \end{equation}
   Using~\eqref{eq:proofProjClosed} we have $\{y^\top R_T:\, y\in\sC\cap\sC^0\}=\{\ty^\top \tR_T:\, \ty\in\tsC\cap\tsC^0\}$
   and now the result follows from Lemma~\ref{le:closednessProbab}, for the special case $\Lambda=\Lambda_i$.

  (ii)~~In the general case, we have $\tsC\cap\tsC^0=(\Lambda_d^\top)^{-1}\circ \cdots \circ (\Lambda_1^\top)^{-1} (\sC\cap\sC^0)$.
  We apply part~(i) successively to $\Lambda_1,\dots,\Lambda_d$ to obtain the result, here~\eqref{eq:proofProjClosed2}
  ensures that the density assumption is satisfied in each step.
\end{proof}

\begin{Lemma}\label{le:pIntegrabilityForLevy}
  Let $p\in (0,1)$ and assume $e_j\in \sC\cap \sC^{0,*}$. Then $u(x_0)<\infty$ implies
  $\int_{\{|x|>1\}} |x^j|^p\,F^R(dx)<\infty$.
\end{Lemma}

\begin{proof}
  Note that $e_j\in \sC^{0,*}$ implies $\Delta R^j>-1$, i.e., $\int_{\{|x|>1\}} |x^j|^p\,F^R(dx)=\int_{\{|x|>1\}} ((x^j)^+)^p\,F^R(dx)$.
  Moreover, we have $E\big[x_0^p\cE(R^j)_T^p\big]\leq u(x_0)$.
  There exists a L\'evy process $Z$ such that
  $\cE(R^j)^p=e^{Z}$, hence $E\big[\cE(R^j)_T^p\big]<\infty$ implies that $\cE(R^j)^p$ is of class (D)
  (cf.~\cite[Lemma 4.4]{Kallsen.00}). In particular, $\cE(R^j)^p$ has a Doob-Meyer decomposition with a well defined drift (predictable finite variation part).
  The stochastic logarithm $Y$ of $\cE(R^j)^p$ is given by $Y=\cE(R^j)^{-p}_-\sint \cE(R^j)^p$ and the drift of $Y$
  is again well defined because the integrand $\cE(R^j)^{-p}_-$ is locally bounded. It\^o's formula shows that $Y$ is a L\'evy process with drift
  \[
    A^Y_t =t \bigg( p (b^R)^j + \tfrac{p(p-1)}{2}(c^R)^{jj} + \int_{\R^d} \big\{(1+x^j)^p-1-p e_j^\top h(x)\big\} \, F^R(dx)\bigg).
  \]
  In particular, $\int_{\{|x|>1\}} ((x^j)^+)^p\,F^R(dx)<\infty$.
\end{proof}

Note that the previous lemma shows Proposition~\ref{pr:levyFinitenessPrelim}(ii); moreover, in the general case, we obtain the desired integrability in the transformed model.

\begin{Cor}
  $u(x_0)<\infty$ implies $\int_{\{|x|>1\}} |x|^p \,F^\tR(dx)<\infty$.
\end{Cor}

\begin{proof}
  By Theorem~\ref{th:levyChangeCoords}(ii) and Corollary~\ref{co:levyChangeCoords}(i) we can apply Lemma~\ref{le:pIntegrabilityForLevy} in the model $(\tR,\tsC)$.
\end{proof}

\begin{Remark}
  The transformation in Theorem~~\ref{th:levyChangeCoords} preserves the L\'evy structure.
  Theorem~\ref{th:levyChangeCoords} and Lemma~\ref{le:pIntegrabilityForLevy} were generalized to semimartingale models in~\cite[Appendix~B]{Nutz.09b}. There, additional assumptions are required for measurable selections and particular structures
  of the model are not preserved in general.
\end{Remark}

\section{Proof of Theorem~\ref{th:MainResultConvex}}\label{se:proofMainThm}

Our aim is to prove Theorem~\ref{th:MainResultConvex} and Proposition~\ref{pr:levyFinitenessPrelim}(i). We shall see that we may replace Assumption~\ref{as:convex}(iv) by the integrability condition~\eqref{eq:LevyPthMoment}. Under~\eqref{eq:LevyPthMoment}, we will obtain the fact that $u(x_0)<\infty$ as we construct the optimal strategies, and that will yield the proof for both results.

\subsection{Solution of the Bellman Equation}\label{se:SolBEllman}

We start with informal considerations to derive a candidate solution for the Bellman equation. It is convenient to start from the general equation for semimartingale models, since then we will be able to apply the verification theory from~\cite{Nutz.09b} without having to argue that our equation is indeed connected to the optimization problem. Moreover, this gives some insight into why the optimal strategy is deterministic in the present L\'evy case.  \emph{A posteriori}, our equation will of course coincide with the one obtained from the standard (formal) dynamic programming equation in the Markovian sense by using the $p$-homogeneity of the value function.

If there is an optimal strategy, \cite[Theorem~3.2]{Nutz.09b} states that the drift rate $a^L$ of the opportunity process $L$ (a special semimartingale in general) satisfies the general Bellman equation
\begin{equation}\label{eq:LDriftRate}
  a^L\,dt = \delta(p-1)L_-^{p/(p-1)}\,dt - p\max_{y\in \sC\cap\sC^0} g(y)\,dt;\quad L_T=1,
\end{equation}
where $g$ is the following function, stated in terms of the joint differential semimartingale characteristics of $(R,L)$:
\begin{align*}%
   g(y)
     & = L_{-}y^\top \Big( b^R + \tfrac{c^{RL}}{L_{-}}+ \tfrac{(p-1)}{2} c^R y \Big) + \int_{\R^d\times\R} x' y^\top h(x)\,F^{R,L}(d(x,x')) \nonumber\\
     &\phantom{:=} + \int_{\R^d\times\R} (L_-+x') \big\{p^{-1}(1+y^\top x)^p - p^{-1} - y^\top h(x)\big\}\, F^{R,L}(d(x,x')).
\end{align*}
In this equation one should see the characteristics of $R$ as the driving terms and $L$ as the solution. In the present L\'evy case, the differential characteristics of $R$ are given by the L\'evy triplet; in particular, they are deterministic (see \cite[II.4.19]{JacodShiryaev.03}). To wit, there is
\emph{no exogenous stochasticity} in~\eqref{eq:LDriftRate}. Therefore we can expect that the opportunity process is deterministic. We make a smooth \emph{Ansatz} $\ell$ for $L$. As $\ell$ has no jumps and vanishing martingale part,
$g$ reduces to $g(y)=\ell\mathfrak{g}(y)$, where $\fg$ is as~\eqref{eq:frakg}.
We show later that this deterministic function is well defined. For the maximization in~\eqref{eq:LDriftRate},
we have the following result.

\begin{Lemma}\label{le:levyMaximizerConstr}
  Suppose that Assumptions~\ref{as:convex}(i)-(iv) hold,
  or alternatively that Assumptions~\ref{as:convex}(i)-(iii) and~\eqref{eq:LevyPthMoment} hold.
  Then $\fg^*:=\sup_{\sC\cap\sC^0} \fg<\infty$ and there exists a vector $\cpi\in \sC\cap\sC^{0,*}$
  such that  $\mathfrak{g}( \cpi)=\mathfrak{g}^*$.
\end{Lemma}

The proof is given in the subsequent section.
As $\ell$ is a smooth function, its drift rate is simply the time derivative, hence we deduce from~\eqref{eq:LDriftRate} that
\[
  d\ell_t = \delta(p-1)\ell_t^{p/(p-1)}\,dt -p \fg^*\ell_t\,dt;\quad \ell_T=1.
\]
This is a Bernoulli ODE.
If we denote $\beta:=1/(1-p)$, the transformation $f(t):=\ell_{T-t}^\beta$ produces the forward linear equation
\[
  \tfrac{d}{dt} f(t)=\delta + \big(\tfrac{p}{1-p}\fg^*\big) f(t);\quad f(0)=1,
\]
which has, with $a=\tfrac{p}{1-p}\fg^*$, the unique solution
$f(t)=-\delta/a + (1+\delta/a)e^{at}$.
Therefore,
\[
  \ell_t=\begin{cases}
    e^{(a/\beta) (T-t)}=e^{p \fg^* (T-t)} & \text{if } \delta=0, \\
    a^{-1/\beta}\big((1+a)e^{a (T-t)} - 1\big)^{1/\beta} & \text{if } \delta=1.
  \end{cases}
\]
If we define $\ckappa_t:=\ell_t^{-\beta}=a\big((1+a)e^{a (T-t)} - 1\big)^{-1}$, then
$(\ell,\cpi,\ckappa)$ is a solution of the Bellman equation in the sense of~\cite[Definition~4.1]{Nutz.09b}.
Let also $\cX=X(\cpi,\ckappa)$ be the corresponding wealth process.
We want to \emph{verify} this solution, i.e., to show that $\ell$ is the opportunity process and that $(\cpi,\ckappa)$ is optimal.
We shall use the following result; it is a special case of Proposition~4.7 and Theorem~5.15 in~\cite{Nutz.09b}.

\begin{Lemma}\label{le:verification}
    The process
     \begin{equation}\label{eq:defGamma}
         \Gamma:=\ell \cX^p+\delta\int \ckappa_s\ell_s \cX^p_s\,ds
     \end{equation}
     is a local martingale. If $\sC$ is convex, then $\Gamma$ is a martingale if and only if $u(x_0)<\infty$ and $(\cpi,\ckappa)$ is optimal and $\ell$ is the opportunity process.
\end{Lemma}

To check that $\Gamma$ is a martingale, it is convenient to consider the closely related process
\[
  \Psi:= p\cpi^\top R^c +\big\{(1+\cpi^\top x)^p - 1\big\}\,\ast(\mu^{R}-\nu^{R}).
\]
Then \cite[Remark~5.9]{Nutz.09b} shows that $\cE(\Psi)$ is a positive local martingale and that $\Gamma$ is a martingale as soon as $\cE(\Psi)$ is. This comes from the fact that $\cE(\Psi)$ equals $\Gamma$ up to a constant if $\delta=0$, while in the case with consumption, the integral in~\eqref{eq:defGamma} is increasing and integrable.
Now $\Psi$ has an advantageous structure:
as $\cpi$ is constant, $\Psi$ is a semimartingale with constant characteristics. In other words,
$\cE(\Psi)$ is an exponential L\'evy local martingale, therefore automatically a true martingale
(e.g.,~\cite[Lemmata~4.2,\,4.4]{Kallsen.00}). Hence we can apply Lemma~\ref{le:verification} to finish the proofs of Theorem~\ref{th:MainResultConvex} and Proposition~\ref{pr:levyFinitenessPrelim}(i), modulo Lemma~\ref{le:levyMaximizerConstr}.

\subsection{Proof of Lemma~\ref{le:levyMaximizerConstr}: Construction of the Maximizer}

Our goal is to show Lemma~\ref{le:levyMaximizerConstr}.
In this section we will use that $\sC$ is star-shaped, but not its convexity. For convenience, we state again the definition
\begin{equation}\label{eq:frakgRestated}
   \fg(y)=y^\top b^R + \tfrac{(p-1)}{2} y^\top c^R y
            + \int_{\R^d} \big\{p^{-1}(1+y^\top x)^p - p^{-1} - y^\top h(x)\big\}\, F^R(dx).
 \end{equation}
The following lemma is a direct consequence of this formula and does not depend on Assumptions~\ref{as:convex}.

\begin{Lemma}\label{le:gProperties}
  \begin{enumerate}[topsep=3pt, partopsep=0pt, itemsep=1pt,parsep=2pt]
    \item If $p\in (0,1)$, $\fg$ is well defined with values in $(-\infty,\infty]$ and lower semicontinuous on $\sC^0$. If~\eqref{eq:LevyPthMoment} holds, $\fg$ is finite and continuous on $\sC^0$.
    \item If $p<0$, $\fg$ is well defined with values in $[-\infty,\infty)$ and upper semicontinuous on $\sC^0$. Moreover,
    $\fg$ is finite on $\check{\sC}$ and $\fg(y)=-\infty$ for $y\in\sC^0\setminus\sC^{0,*}$.
  \end{enumerate}
\end{Lemma}

\begin{proof}
  Fix a sequence $y_n\to y$ in $\sC^0$. A Taylor expansion and~\eqref{eq:LevyIntegrates} show that
  $\int_{|x|\leq \eps} \big\{p^{-1}(1+y^\top x)^p - p^{-1} - y^\top h(x)\big\}\, F^R(dx)$ is finite
  and continuous along $(y_n)$ for $\eps$ small enough, e.g., $\eps= (2 \sup_n |y_n|)^{-1}$.

  If $p<0$, we have $p^{-1}(1+y^\top x)^p - p^{-1} - y^\top h(x)\leq - p^{-1} - y^\top h(x)$ and Fatou's lemma shows that
  $\int_{|x|> \eps} \big\{p^{-1}(1+y^\top x)^p - p^{-1} - y^\top h(x)\big\}\, F^R(dx)$ is upper semicontinuous of with respect to $y$. For $p>0$ we have the converse inequality and the same argument yields
  lower semicontinuity. If $p>0$ and~\eqref{eq:LevyPthMoment} holds, then as the integrand grows at most like $|x|^p$ at infinity, the integral is finite and dominated convergence yields continuity.

  Let $p<0$. For finiteness on $\check{\sC}$ we note that $\fg$ is even finite on $\lambda \sC^0$ for any $\lambda\in [0,1)$. Indeed, $y\in \lambda \sC^0$ implies $y^\top x\geq -\lambda >-1$ $F^R(dx)$-a.e., hence the integrand in~\eqref{eq:frakgRestated} is bounded $F^R$-a.e.\ and we
  conclude by~\eqref{eq:LevyIntegrates}. The last claim is immediate from the definitions of $\sC^0$ and $\sC^{0,*}$ as well as~\eqref{eq:frakgRestated}.
\end{proof}%

Assume the version of Lemma~\ref{le:levyMaximizerConstr} under Assumptions~\ref{as:convex}(i)-(iii) and~\eqref{eq:LevyPthMoment}
has already been proved; we argue that the complete claim of that lemma then follows. Indeed, suppose that Assumptions~\ref{as:convex}(i)-(iv) hold.
We first observe that $\sC\cap\sC^{0,*}$ is dense in $\sC\cap\sC^0$. To see this, note that
for $y\in\sC\cap\sC^0\setminus\sC^{0,*}$ and $n\in\N$ we have $y_n:=(1-n^{-1})y\to y$ and $y_n$ is in $\sC^{0,*}$ (by the definition)
and also in $\sC$, due to the star-shape.
Using Section~\ref{se:transformation} and its notation, Assumptions~\ref{as:convex}(i)-(iv) now imply that the transformed model $(\tR,\tsC)$ satisfies
Assumptions~\ref{as:convex}(i)-(iii) and~\eqref{eq:LevyPthMoment}. We apply our version of
Lemma~\ref{le:levyMaximizerConstr} in that model to
obtain $\tilde{\fg}^*:=\sup_{\tsC\cap\tsC^0} \tilde{\fg}<\infty$ and a vector $\tpi\in \tsC\cap\tsC^{0,*}$
such that $\tilde{\fg}(\tpi)=\tilde{\fg}^*$.
The density of $\sC\cap\sC^{0,*}$ observed above and the semicontinuity from Lemma~\ref{le:gProperties}(i) imply
that $\fg^*:=\sup_{\sC\cap\sC^0} \fg=\sup_{\sC\cap\sC^{0,*}} \fg$. Using this argument also for $(\tR,\tsC)$,
Corollary~\ref{co:levyChangeCoords}(iii) yields $\fg^*=\tilde{\fg}^*<\infty$. Moreover,
Corollary~\ref{co:levyChangeCoords}(iv) states that $\cpi:=\Lambda^\top \tpi\in \sC\cap\sC^{0,*}$ is a maximizer for $\fg$.

Summarizing this discussion, it suffices to prove Lemma~\ref{le:levyMaximizerConstr} under Assumptions~\ref{as:convex}(i)-(iii) and~\eqref{eq:LevyPthMoment}; hence these will be our assumptions for the rest of the section.

Formally, by differentiation under the integral, the directional derivatives of $\fg$ are given by
$(\ty-y)^\top \nabla \mathfrak{g}(y)=\mathfrak{G}(\ty,y)$, with
\begin{align}\label{eq:frakG}
  \fG(\ty,y)&:= (\ty-y)^\top \big( b^R + (p-1)c^R y \big)\\
     & \phantom{:=} + \int_{\R^d} \Big\{\frac{(\ty-y)^\top x}{(1+y^\top x)^{1-p}} - (\ty-y)^\top h(x)\Big\}\, F^R(dx).\nonumber
\end{align}
We take this as the definition of $\fG(\ty,y)$ whenever the integral makes sense.

\begin{Remark}
 Formally setting $p=0$, we see that $\fG$
 corresponds to the \emph{relative rate of return} of two portfolios in the theory of $\log$-utility \cite[Eq.~(3.2)]{KaratzasKardaras.07}.
\end{Remark}

\begin{Lemma}\label{le:GProperties}
  Let $\ty\in\sC^0$. On the set $\sC^0\cap \{\fg>-\infty\}$, $\fG(\ty,\cdot)$ is well defined with values in $(-\infty,\infty]$.
  Moreover, $\fG(0,\cdot)$ is lower semicontinuous on $\sC^0\cap \{\fg>-\infty\}$. %
\end{Lemma}

\begin{proof}
  The first part follows by rewriting $\fG(\ty,y)$ as
  \begin{align*}
  (\ty-y)^\top &\big( b^R + (p-1)c^R y \big) - \int \big\{(1+y^\top x)^p -1 -p y^\top h(x)\big\}\, F^R(dx) \nonumber\\
     &+ \int \Big\{\frac{1+\ty^\top x}{(1+y^\top x)^{1-p}} -1 -(\ty+(p-1)y)^\top h(x)\Big\}\, F^R(dx)
  \end{align*}
  because the first integral occurs in~\eqref{eq:frakgRestated} and $1+\ty^\top x\geq0$ $F^R$-a.e.\ by definition of $\sC^0$.
  Let $p\in(0,1)$ and $\ty=0$ in the definition of $\fG$. Using
  \[
   \frac{-y^\top x}{(1+y^\top x)^{1-p}} \geq -\frac{1+y^\top x}{(1+y^\top x)^{1-p}}=-(1+y^\top x)^{p}
  \]
  and~\eqref{eq:LevyPthMoment}, Fatou's lemma yields that $\fG(0,\cdot)$ is l.s.c.\ on $\sC^0$.
  If $p<0$, then $z/(1+z)^{1-p}\leq 1$ for $z\geq 0$ %
  implies
  $\frac{-y^\top x}{(1+y^\top x)^{1-p}}\geq -1$. Again, Fatou's lemma yields the claim.
\end{proof}
As our goal is to prove Lemma~\ref{le:levyMaximizerConstr}, we may assume in the following  that
\[
 \sC\cap\sC^0\subseteq \sN^\bot.
\]
Indeed, noting that $\fg(y)=\fg(y+y')$ for $y'\in\sN$, we may
substitute $\sC\cap\sC^0$ by its projection to $\sN^\bot$.
The remainder of the section parallels the case of $\log$-utility as treated in~\cite[Lemmata 5.2,5.1]{Kardaras.09}.
By Lemmata~\ref{le:gProperties} and~\ref{le:GProperties}, $\fG(0,y)$ is well defined with values in $(-\infty,\infty]$ for $y\in \check{\sC}$, so the following statement makes sense.

\begin{Lemma}\label{le:NUIPandDerivative}
  Let $y\in (\sC\cap\sC^0)^{\check{}}$, then $y\in\sJ$ if and only if $\fG(0,ay)\leq 0$ for all $a\geq0$.
\end{Lemma}

\begin{proof}
  If $y\in\sJ$, then $\fG(0,ay)\leq 0$ by the definitions of $\sJ$ and $\fG$; we prove the converse.
  As $y\in (\sC\cap\sC^0)^{\check{}}$ implies $F^R[ay^\top x< -1]=0$ for all $a$, we have $F^R[y^\top x<0]=0$.
  Since $y^\top x\geq0$ entails $|\frac{y^\top x}{(1+y^\top x)^{1-p}}|\leq 1$, dominated convergence yields
  \[
    \lim_{a\to\infty} \int \Big\{\frac{y^\top x}{(1+ay^\top x)^{1-p}} - y^\top h(x)\Big\}\, F^R(dx)=-\int y^\top h(x)\, F^R(dx).
  \]
  By assumption, $-a^{-1}\fG(0,ay)\geq0$, i.e,
  \[
   y^\top b^R + a(p-1)y^\top c^R y
     + \int \Big\{\frac{y^\top x}{(1+ay^\top x)^{1-p}} - y^\top h(x)\Big\}\, F^R(dx)\geq0.
  \]
 As $(p-1)y^\top c^R y\leq0$, taking $a\to\infty$ shows $y^\top c^R =0$ and then we also see $y^\top b^R -\int y^\top h(x)\,F(dx)\geq0$.
\end{proof}

\begin{proof}[Proof of Lemma~\ref{le:levyMaximizerConstr}]
  Let $(y_n)\subset \sC\cap\sC^0$ be such that $\mathfrak{g}(y_n)\to \mathfrak{g}^*$.
   We may assume $\fg(y_n)>-\infty$ and $y_n\in\sC^{0,*}$ by Lemma~\ref{le:gProperties}, since $\sC^{0,*}\subseteq \sC^0$ is dense.

  We claim that $(y_n)$ has a bounded subsequence. By way of contradiction, suppose that $(y_n)$ is unbounded. Without loss of generality, $\xi_n:=y_n/|y_n|$ converges to some $\xi$. Moreover, we may assume by redefining $y_n$ that $\mathfrak{g}(y_n)=\max_{\lambda \in [0,1]} \mathfrak{g}(\lambda  y_n)$, because
  $\mathfrak{g}$ is continuous on each of the compact sets $C_n=\{\lambda  y_n:\, \lambda\in [0,1]\}$. Indeed, if $p<0$, continuity follows by dominated convergence using $1+\lambda y^\top x \geq 1+ y^\top x$ on $\{x:\,y^\top x\leq0\}$; while for $p\in(0,1)$, $\fg$ is continuous by Lemma~\ref{le:gProperties}.

  Using concavity one can check that $\fG(0,a\xi_n)$ is indeed the directional derivative of the function $\fg$ at $a\xi_n$ (cf.~\cite[Lemma~5.14]{Nutz.09b}). In particular, $\mathfrak{g}(y_n)=\max_{\lambda \in [0,1]} \mathfrak{g}(\lambda  y_n)$ implies that $\fG(0,a\xi_n)\leq 0$ for $a>0$ and all $n$ such that $|y_n|\geq a$ (and hence $a\xi_n\in C_n$).
  By the star-shape and closedness of $\sC\cap\sC^0$ we have $\xi\in(\sC\cap\sC^0)^{\check{}}$. Lemmata~\ref{le:gProperties} and \ref{le:GProperties} yield the semicontinuity to pass from $\fG(0,a\xi_n)\leq 0$ to $\fG(0,a\xi)\leq 0$ and now Lemma~\ref{le:NUIPandDerivative} shows $\xi\in\sJ$, contradicting the $\mathrm{NUIP}_{\sC}$ condition that $\sJ\cap(\sC\cap\sC^0)^{\check{}}=\emptyset$.

  We have shown that after passing to a subsequence, there exists a limit $y^*=\lim_n y_n$.
  Lemma~\ref{le:gProperties} shows $\mathfrak{g}^*=\lim_n \mathfrak{g}(y_n)=\mathfrak{g}(y^*)<\infty$; and
  $y^*\in\sC^{0,*}$ for $p<0$. For $p\in (0,1)$, $y^*\in\sC^{0,*}$ follows as in~\cite[Lemma~A.8]{Nutz.09b}.
\end{proof}

\section{$q$-Optimal Martingale Measures}\label{se:qOptMeas}

In this section we consider $\delta=0$ (no consumption) and $\sC=\R^d$. Then Assumptions~\ref{as:convex}
are equivalent to
\begin{equation}\label{eq:AssumptionsqOpt}
  \mathrm{NUIP_{\R^d}} \mbox{ holds\quad and}\quad u(x_0)<\infty
\end{equation}
and these conditions are in force for the following discussion. Let
$\sM$ be the set of all equivalent local martingale measures for $S=\cE(R)$. Then $\mathrm{NUIP_{\R^d}}$ is equivalent to
$\sM\neq\emptyset$, more precisely, there exists $Q\in\sM$ under which $R$ is a L\'evy martingale (see~\cite[Remark~3.8]{Kardaras.09}). In particular, we are in the setting of Kramkov and Schachermayer~\cite{KramkovSchachermayer.99}.

Let $q=p/(p-1)\in (-\infty,0)\cup (0,1)$ be the exponent conjugate to $p$,
then $Q\in\sM$ is called \emph{$q$-optimal} if $E[-q^{-1}(dQ/dP)^q]$ is finite and minimal over $\sM$. If $q<0$, i.e., $p\in (0,1)$,
then $u(x_0)<\infty$ is equivalent to the existence of some $Q\in\sM$ such that $E[-q^{-1}(dQ/dP)^q]<\infty$ (cf.\ Kramkov and Schachermayer~\cite{KramkovSchachermayer.03}).

This minimization problem over $\sM$ is linked to power utility maximization by convex duality in the sense of~\cite{KramkovSchachermayer.99}. More precisely, that article considers a ``dual problem'' over an enlarged domain of certain supermartingales.
We recall from \cite[Proposition~4.2]{Nutz.09a} that the solution to that dual problem is given by the positive supermartingale $\hY=L\hX^{p-1}$,
where $L$ is the opportunity process and $\hX=x_0\cE(\hpi\sint R)$ is the optimal wealth process corresponding to $\hpi$ as in Theorem~\ref{th:MainResultConvex}.
It follows from~\cite[Theorem~2.2(iv)]{KramkovSchachermayer.99} that the $q$-optimal martingale measure $\widehat{Q}$ exists if and only if $\hY$ is a martingale, and in that case $\hY/\hY_0$ is the $P$-density process of $\widehat{Q}$.
Recall the functions $\fg$ and $\fG$ from~\eqref{eq:frakg} and~\eqref{eq:frakG}. A direct calculation (or \cite[Remark~5.18]{Nutz.09b}) shows
\begin{align*}%
  \hY/\hY_0= \cE\Big(&\fG(0,\hpi) t + (p-1)\hpi^\top R^c
              +\big\{(1+\hpi^\top x)^{p-1} - 1\}\,\ast(\mu^{R}-\nu^{R})\Big).
\end{align*}
Here absence of drift is equivalent to $\fG(0,\hpi)=0$, or more explicitly,
\begin{equation}\label{eq:drift0}
  \hpi^\top b^R + (p-1)\hpi^\top c^R \hpi
    + \int_{\R^d} \Big\{\frac{ \hpi^\top x}{(1+\hpi^\top x)^{1-p}} - \hpi^\top h(x)\Big\}\, F^R(dx)=0,
\end{equation}
and in that case
\begin{equation}\label{eq:optDensity}
  \hY/\hY_0= \cE\Big( (p-1)\hpi^\top R^c +\big\{(1+\hpi^\top x)^{p-1} - 1 \big\}\,\ast(\mu^{R}-\nu^{R})\Big).
\end{equation}
This is an exponential L\'evy martingale because $\hpi$ is a constant vector; in particular, it is indeed a true martingale.

\begin{Thm}\label{th:qOptMeas}
  The following are equivalent:
  \begin{enumerate}[topsep=3pt, partopsep=0pt, itemsep=1pt,parsep=2pt]
    \item The $q$-optimal martingale measure $\widehat{Q}$ exists,
    \item \eqref{eq:AssumptionsqOpt} and~\eqref{eq:drift0} hold,
    \item there exists $\hpi\in\sC^0$ such that $\fg(\hpi)=\max_{\sC^0} \fg<\infty$ and~\eqref{eq:drift0} holds.
  \end{enumerate}
  Under these equivalent conditions,~\eqref{eq:optDensity} is the $P$-density process of $\widehat{Q}$.
\end{Thm}

\begin{proof}
  We have just argued the equivalence of (i) and (ii). Under~\eqref{eq:AssumptionsqOpt}, there exists $\hpi$ satisfying (iii) by Theorem~\ref{th:MainResultConvex}. Conversely, given (iii) we construct
  the solution to the utility maximization problem as before and~\eqref{eq:AssumptionsqOpt} follows; recall Remark~\ref{rk:onAssumptions}(c).
\end{proof}

\begin{Remark}\label{re:qOptMeas}
  (i)~~If $\widehat{Q}$ exists,~\eqref{eq:optDensity} shows that the change of measure from $P$ to $\widehat{Q}$ has constant (deterministic and time-independent) Girsanov parameters $\big((p-1)\hpi,(1+\hpi^\top x)^{p-1}\big)$; compare~\cite[III.3.24]{JacodShiryaev.03} or Jeanblanc et al.~\cite[\S A.1, \S A.2]{JeanblancKloppelMiyahara.07}.
  Therefore, \emph{$R$ is again a L\'evy process under $\widehat{Q}$}. This result was previously obtained in~\cite{JeanblancKloppelMiyahara.07} by an abstract argument (cf.\ Section~\ref{se:literature} below).

  (ii)~~The existence of $\widehat{Q}$ is a fairly delicate question compared to the existence of the supermartingale $\hY$. Recalling the definition~\eqref{eq:frakG} of $\fG$,~\eqref{eq:drift0} essentially expresses that the budget constraint $\sC^0$
  in the maximization of $\fg$ is ``not binding''. Theorem~\ref{th:qOptMeas} gives an explicit and sharp
  description for the existence of $\widehat{Q}$; this appears to be missing in the previous literature.
\end{Remark}

\section{Extensions to Non-Convex Constraints}\label{se:nonconvexConstraints}
In this section we consider the utility maximization problem for some cases where the constraints $0\in\sC\subseteq \R^d$ are not convex.
Let us first recapitulate where the convexity assumption was
used above. The proof of Lemma~\ref{le:levyMaximizerConstr} used
the star-shape of $\sC$, but not convexity. In the rest of Section~\ref{se:SolBEllman}, the shape of $\sC$
was irrelevant except in Lemma~\ref{le:verification}.

We denote by $\csC$ the closed convex hull of $\sC$.

\begin{Cor}\label{co:nonconvexPneg}
  Let $p<0$ and suppose that either (i) or (ii) below hold:
  \begin{enumerate}[topsep=3pt, partopsep=0pt, itemsep=1pt,parsep=2pt]
    \item
        \begin{enumerate}
          \item $\sC$ is star-shaped,
          \item the orthogonal projection of $\csC\cap\sC^0$ onto $\sN^\bot$ is closed,
          \item $\mathrm{NUIP}_{\csC}$ holds.
        \end{enumerate}
        \item $\sC\cap\sC^0$ is compact.
  \end{enumerate}
  Then the assertion of Theorem~\ref{th:MainResultConvex} remains valid.
\end{Cor}

\begin{proof}
  (i)~~The construction of $(\ell,\cpi,\ckappa)$ is as above; we have to substitute the verification step which used Lemma~\ref{le:verification}.
  The model $(R,\csC)$ satisfies the assumptions of Theorem~\ref{th:MainResultConvex}. Hence the corresponding opportunity process
  $L^{\csC}$ is deterministic and bounded away from zero. The definition of the opportunity process and the inclusion
  $\sC\subseteq\csC$ imply that  the opportunity process
  $L=L^\sC$ for $(R,\sC)$ is also bounded away from zero. Hence $\ell/L$ is bounded and we can verify $(\ell,\cpi,\ckappa)$ by
  \cite[Corollary~5.4]{Nutz.09b}, which makes no assumptions about the shape of $\sC$.

  (ii) We may assume without loss of generality that $\sC=\sC\cap\sC^0$. In (i), the star-shape was used only to construct a maximizer for $\fg$.
  When $\sC\cap\sC^0$ is compact, its existence is clear by the upper semicontinuity from Lemma~\ref{le:gProperties}, which also shows that any maximizer is necessarily in $\sC^{0,*}$. To proceed as in (i), it remains to note that
  the projection of the compact set $\csC \cap\sC^0$ onto $\sN^\bot$ is compact, and $\mathrm{NUIP}_{\csC}$ holds because
  $(\csC)^{\check{}}=\{0\}$ since $\csC$ is bounded.
\end{proof}

When the constraints are not star-shaped and $p>0$, an additional condition is necessary to ensure that
the maximum of $\fg$ is not attained on $\sC^0\setminus \sC^{0,*}$, or equivalently, to obtain a positive optimal
wealth process. In~\cite[\S 2.4]{Nutz.09b}, the following condition was introduced:
\begin{description}
 \item[(C3)] There exists $\underline{\eta}\in (0,1)$ such that
            $y\in (\sC\cap \sC^{0}) \setminus \sC^{0,*}$ implies $\eta y \in \sC$ for all $\eta\in (\underline{\eta},1)$.
\end{description}
This is clearly satisfied if $\sC$ is star-shaped or if $\sC^{0,*}=\sC^0$.

\begin{Cor}\label{co:nonConvexPpos}
  Let $p\in (0,1)$ and suppose that either (i) or (ii) below hold:
  \begin{enumerate}[topsep=3pt, partopsep=0pt, itemsep=1pt,parsep=2pt]
    \item Assumptions~\ref{as:convex} hold except that $\sC$ is star-shaped instead of being convex.
    \item $\sC\cap\sC^0$ is compact and satisfies (C3) and $u(x_0)<\infty$.
  \end{enumerate}
  Then the assertion of Theorem~\ref{th:MainResultConvex} remains valid.
\end{Cor}

\begin{proof}
  (i)~~The assumptions carry over to the transformed model as before, hence again we only need to substitute the verification argument. In view of $p\in(0,1)$, we can use \cite[Theorem~5.2]{Nutz.09b}, which makes no assumptions about the shape of $\sC$. Note that we have already checked its condition (cf.\ \cite[Remark~5.16]{Nutz.09b}).

  (ii)~~We may again assume $\sC=\sC\cap\sC^0$ and Remark~\ref{rk:shapePreservation} shows that we can choose $\tsC\cap\tsC^0$ to be compact in the transformed model satisfying~\eqref{eq:LevyPthMoment}. That is, we can again assume~\eqref{eq:LevyPthMoment} without loss of generality.
  Then $\fg$ is continuous and hence existence of a maximizer on $\sC\cap\sC^0$ is clear. Under~(C3), any maximizer is in $\sC^{0,*}$ by the same argument as in the proof of Lemma~\ref{le:levyMaximizerConstr}.
\end{proof}

The following result covers \emph{all} closed constraints and applies to most of the standard models (cf.\ Example~\ref{ex:simpleIllustration}).

\begin{Cor}
  Let $\sC$ be closed and assume that $\sC^0$ is compact and that $u(x_0)<\infty$. Then the assertion of Theorem~\ref{th:MainResultConvex} remains valid.
\end{Cor}
\begin{proof}
  Note that (C3) holds for all sets $\sC$ when $\sC^{0,*}$ is closed (and hence equal to $\sC^0$). It remains to apply part~(ii) of the two previous corollaries.
\end{proof}

\begin{Remark}
  (i)~~For $p\in(0,1)$ we also have the analogue of Proposition~\ref{pr:levyFinitenessPrelim}(i): under the assumptions of Corollary~\ref{co:nonConvexPpos} excluding $u(x_0)<\infty$, $\eqref{eq:LevyPthMoment}$ implies $u(x_0)<\infty$.

 (ii)~~ The optimal propensity to consume $\hkappa$ remains unique even when the constraints are not convex (cf.\ \cite[Theorem~3.2]{Nutz.09b}). However, there is no uniqueness for the optimal portfolio. In fact, in the setting of the above corollaries, any constant vector $\pi\in\argmax_{\sC\cap\sC^0} \fg$ is an optimal portfolio (by the same proofs); and when $\sC$ is not convex, the difference of two such $\pi$ need not be in $\sN$. See also~\cite[Remark~3.3]{Nutz.09b} for statements about dynamic portfolios. Conversely,
 by \cite[Theorem~3.2]{Nutz.09b} any optimal portfolio, possibly dynamic, takes values in $\argmax_{\sC\cap\sC^0}\fg$.
\end{Remark}

\section{Related Literature}\label{se:literature}

We discuss some related literature in a highly selective manner; an exhaustive overview is beyond our scope.
For the unconstrained utility maximization problem with general L\'evy processes, Kallsen~\cite{Kallsen.00} gave a result of verification type: If there exists a vector $\pi$ satisfying a certain equation, $\pi$ is the optimal portfolio. This equation is essentially our~\eqref{eq:drift0} and therefore holds only if the corresponding dual element $\widehat{Y}$ is the density process of a measure.
Muhle-Karbe~\cite[Example~4.24]{MuhleKarbe.09} showed that this condition fails in a particular model. In the one-dimensional case, he introduced a weaker inequality condition~\cite[Corollary 4.21]{MuhleKarbe.09}, but again existence of $\pi$ was not discussed.
(In fact, our proofs show the necessity of that inequality condition; cf.~\cite[Remark~5.16]{Nutz.09b}.)

Numerous variants of our utility maximization problem were also studied along more traditional lines of dynamic programming. E.g.,
Benth et al.~\cite{BenthKarlsenReikvam.01} solve a similar problem with infinite time horizon when the L\'evy process satisfies additional integrability properties and the portfolios are chosen in the interval $[0,1]$. This part of the literature generally requires technical conditions, which we sought to avoid. %

Jeanblanc et al.~\cite{JeanblancKloppelMiyahara.07} study the $q$-optimal measure $\widehat{Q}$ for L\'evy processes when $q<0$ or $q>1$ (note that the considered parameter range does not coincide with ours). They show that the L\'evy structure is preserved under $\widehat{Q}$, if the latter exists; a result we recovered in Remark~\ref{re:qOptMeas} above for our values of $q$.
In~\cite{JeanblancKloppelMiyahara.07} this is established by showing that starting from any equivalent change of measure, a suitable choice of constant Girsanov parameters reduces the $q$-divergence of the density. This argument does not seem to extend to our general dual problem which involves supermartingales rather than measures; in particular, it cannot be used to show that the optimal portfolio is a constant vector.
A deterministic, but not explicit characterization of $\widehat{Q}$ is given in~\cite[Theorem~2.7]{JeanblancKloppelMiyahara.07}. The authors also provide a more explicit candidate for
the $q$-optimal measure~\cite[Theorem~2.9]{JeanblancKloppelMiyahara.07}, but the condition of that theorem fails in general (see
Bender and Niethammer~\cite{BenderNiethammer.08}).

In the L\'evy setting the $q$-optimal measures ($q\in\R$) coincide with the
minimal Hellinger measures and hence the pertinent results apply. See Choulli and Stricker~\cite{ChoulliStricker.09} and in particular their general sufficient condition~\cite[Theorem~2.3]{ChoulliStricker.09}. We refer to~\cite[p.\,1623]{JeanblancKloppelMiyahara.07} for a discussion.
Our result differs in that both the existence of $\widehat{Q}$ and its density process are described explicitly in terms of the L\'evy triplet.
\phantom{\cite{CawstonVostrikova.10} \cite{Grandits.00} \cite{ElKaroui.81}}
\newcommand{\dummy}[1]{}


\begin{thebibliography}{10}

\bibitem{BenderNiethammer.08}
C.~Bender and C.~R. Niethammer.
\newblock On {$q$}-optimal martingale measures in exponential {L}{\'e}vy
  models.
\newblock {\em Finance Stoch.}, 12(3):381--410, 2008.

\bibitem{BenthKarlsenReikvam.01}
F.~E. Benth, K.~H. Karlsen, and K.~Reikvam.
\newblock Optimal portfolio management rules in a non-{G}aussian market with
  durability and intertemporal substitution.
\newblock {\em Finance Stoch.}, 5(4):447--467, 2001.

\bibitem{BenthKarlsenReikvam.01b}
F.~E. Benth, K.~H. Karlsen, and K.~Reikvam.
\newblock Optimal portfolio selection with consumption and nonlinear
  integro-differential equations with gradient constraint: a viscosity solution
  approach.
\newblock {\em Finance Stoch.}, 5(3):275--303, 2001.

\bibitem{CawstonVostrikova.10}
S.~Cawston and L.~Vostrikova.
\newblock Minimal {$f$}-divergence martingale measures and optimal portfolios
  for exponential {L}{\'e}vy models with a change-point.
\newblock {\em Preprint arXiv:1004.3525v1}, 2010.

\bibitem{ChoulliStricker.09}
T.~Choulli and C.~Stricker.
\newblock Comparing the minimal {H}ellinger martingale measure of order {$q$}
  to the {$q$}-optimal martingale measure.
\newblock {\em Stochastic Process. Appl.}, 119(4):1368--1385, 2009.

\bibitem{CzichowskySchweizer.09a}
C.~Czichowsky and M.~Schweizer.
\newblock Closedness in the semimartingale topology for spaces of stochastic
  integrals with constrained integrands.
\newblock To appear in {\em S{\'e}m. Probab.}, 2009.

\bibitem{CzichowskyWestrayZheng.08}
C.~Czichowsky, N.~Westray, and H.~Zheng.
\newblock Convergence in the semimartingale topology and constrained
  portfolios.
\newblock To appear in {\em S{\'e}m. Probab.}, 2008.
%
%
%
%
%
%
%

\bibitem{Foldes.90}
L.~Foldes.
\newblock Conditions for optimality in the infinite-horizon
  portfolio-cum-saving problem with semimartingale investments.
\newblock {\em Stochastics Stochastics Rep.}, 29(1):133--170, 1990.

\bibitem{FramstadOksendalSulem.99}
N.~C. Framstad, B.~{\O}ksendal, and A.~Sulem.
\newblock Optimal consumption and portfolio in a jump diffusion market.
\newblock {\em Discussion Paper, Norwegian School of Economics and Business
  Administration}, 1999.

\bibitem{Grandits.00}
P.~Grandits.
\newblock On martingale measures for stochastic processes with independent
  increments.
\newblock {\em Theory Probab. Appl.}, 44(1):39--50, 2000.

\bibitem{HarrisonPliska.81}
J.~M. Harrison and S.~R. Pliska.
\newblock Martingales and stochastic integrals in the theory of continuous
  trading.
\newblock {\em Stochastic Process. Appl.}, 11(3):215--260, 1981.

\bibitem{JacodShiryaev.03}
J.~Jacod and A.~N. Shiryaev.
\newblock {\em Limit Theorems for Stochastic Processes}.
\newblock Springer, Berlin, 2nd edition, 2003.

\bibitem{JeanblancKloppelMiyahara.07}
M.~Jeanblanc, S.~Kl{\"o}ppel, and Y.~Miyahara.
\newblock Minimal {$f^q$}-martingale measures for exponential {L}{\'e}vy
  processes.
\newblock {\em Ann. Appl. Probab.}, 17(5/6):1615--1638, 2007.

\bibitem{Kallsen.00}
J.~Kallsen.
\newblock Optimal portfolios for exponential {L\'{e}}vy processes.
\newblock {\em Math. Methods Oper. Res.}, 51(3):357--374, 2000.

\bibitem{KaratzasKardaras.07}
I.~Karatzas and C.~Kardaras.
\newblock The num\'{e}raire portfolio in semimartingale financial models.
\newblock {\em Finance Stoch.}, 11(4):447--493, 2007.

\bibitem{KaratzasZitkovic.03}
I.~Karatzas and G.~{\v Z}itkovi\'c.
\newblock Optimal consumption from investment and random endowment in
  incomplete semimartingale markets.
\newblock {\em Ann. Probab.}, 31(4):1821--1858, 2003.

\bibitem{Kardaras.09}
C.~Kardaras.
\newblock No-free-lunch equivalences for exponential {L}\'evy models under
  convex constraints on investment.
\newblock {\em Math. Finance}, 19(2):161--187, 2009.

\bibitem{KramkovSchachermayer.99}
D.~Kramkov and W.~Schachermayer.
\newblock The asymptotic elasticity of utility functions and optimal investment
  in incomplete markets.
\newblock {\em Ann. Appl. Probab.}, 9(3):904--950, 1999.

\bibitem{KramkovSchachermayer.03}
D.~Kramkov and W.~Schachermayer.
\newblock Necessary and sufficient conditions in the problem of optimal
  investment in incomplete markets.
\newblock {\em Ann. Appl. Probab.}, 13(4):1504--1516, 2003.

\bibitem{Merton.69}
R.~C. Merton.
\newblock Lifetime portfolio selection under uncertainty: the continuous-time
  case.
\newblock {\em Rev. Econom. Statist.}, 51:247--257, 1969.

\bibitem{Mossin.68}
J.~Mossin.
\newblock Optimal multiperiod portfolio policies.
\newblock {\em J. Bus.}, 41(2):215--229, 1968.

\bibitem{MuhleKarbe.09}
J.~Muhle-Karbe.
\newblock {\em On Utility-Based Investment, Pricing and Hedging in Incomplete
  Markets}.
\newblock PhD thesis, TU M{\"u}nchen, 2009.

\bibitem{Nutz.09b}
M.~Nutz.
\newblock The {B}ellman equation for power utility maximization with
  semimartingales.
\newblock {\em Preprint arXiv:0912.1883v1}, 2009.

\bibitem{Nutz.09a}
M.~Nutz.
\newblock The opportunity process for optimal consumption and investment with
  power utility.
\newblock To appear in {\em Math. Financ. Econ.}, 2009.

\bibitem{Samuelson.69}
P.~A. Samuelson.
\newblock Lifetime portfolio selection by dynamic stochastic programming.
\newblock {\em Rev. Econ. Statist.}, 51(3):239--246, 1969.

\bibitem{Sato.99}
K.-I. Sato.
\newblock {\em L{\'e}vy Processes and Infinitely Divisible Distributions}.
\newblock Cambridge University Press, Cambridge, 1999.

\bibitem{Schoutens.03}
W.~Schoutens.
\newblock {\em L{\'e}vy Processes in Finance: Pricing Financial Derivatives}.
\newblock Wiley, Chichester, 2003.

\end{thebibliography}
\end{document}